\documentclass[conference]{IEEEtran}
\IEEEoverridecommandlockouts
\usepackage{cite}
\usepackage{amsmath,amssymb,amsfonts}
\usepackage{algorithmic}
\usepackage{graphicx}
\usepackage{textcomp}
\usepackage{xcolor}
\def\BibTeX{{\rm B\kern-.05em{\sc i\kern-.025em b}\kern-.08em
    T\kern-.1667em\lower.7ex\hbox{E}\kern-.125emX}}

\let\abs=\envert

\newcommand\opc[1]{{\cal #1}}
\newcommand\un[1]{{\rm\,#1}}

\newcommand\fFC[1]{{\rm F}^{(#1)}_{\rm C}}
\newcommand\fnFC[1]{{\rm\widetilde{F}}^{(#1)}_{\rm C}}

\newcommand\numeq[1]{(\ref{#1})}
\newcommand\eq[1]{Eq.~(\ref{#1})}

\newcommand\fig[1]{Fig.~\ref{#1}}
\newcommand\numfig[1]{\ref{#1}}
\newcommand\figs[1]{Figs.~\ref{#1}}
\newcommand\sect[1]{Section~\ref{#1}}

\newcommand\insertfig[6]
{\begin{figure}[#1]
    \begin{center}
        \includegraphics[bb = 0 0 #4cm #5cm]{#3.jpg}
    \end{center}
    \caption{#6}
    \label{#2}
\end{figure}}

\newcommand\insertfigdc[6]
{\begin{figure*}[#1]
    \begin{center}
        \includegraphics[bb = 0 0 #4cm #5cm]{#3.jpg}
    \end{center}
    \caption{#6}
    \label{#2}
\end{figure*}}

\begin{document}
\pagestyle{plain}

\title{Spectral analysis of signals by time-domain statistical characterization and neural network processing: Application to correction of spectral amplitude alterations in pulse-like waveforms\\
}

\author{\IEEEauthorblockN{1\textsuperscript{st} Guillermo H. Bustos}\\
\IEEEauthorblockA{\textit{FAMAF - UNC, IFEG - CONICET, Córdoba, Argentina}\\
ghbustos@famaf.unc.edu.ar}
\and
\IEEEauthorblockN{2\textsuperscript{nd} Héctor H. Segnorile}\\
\IEEEauthorblockA{\textit{FAMAF - UNC, IFEG - CONICET, Córdoba, Argentina}\\
segnorile@famaf.unc.edu.ar}
}

%\author{\IEEEauthorblockN{1\textsuperscript{st} Guillermo H. Bustos}
%\IEEEauthorblockA{\textit{Facultad de Matemática, Astronomía, Física y Computación (FAMAF) - Universidad Nacional de Córdoba (UNC)} \\
%\textit{Instituto de Física Enrique Gaviola (IFEG) - CONICET}\\
%M. Allende y H. de la Torre - Ciudad Universitaria, X5016LAE - Córdoba, Argentina\\
%ghbustos@famaf.unc.edu.ar}
%\and
%\IEEEauthorblockN{2\textsuperscript{nd} Héctor H. Segnorile}
%\IEEEauthorblockA{\textit{Facultad de Matemática, Astronomía, Física y Computación (FAMAF) - Universidad Nacional de Córdoba (UNC)} \\
%\textit{Instituto de Física Enrique Gaviola (IFEG) - CONICET}\\
%M. Allende y H. de la Torre - Ciudad Universitaria, X5016LAE - Córdoba, Argentina\\
%segnorile@famaf.unc.edu.ar - https://orcid.org/0000-0002-9060-6342}
%}

\maketitle
\thispagestyle{plain}

\begin{abstract}
We present a time-domain method to detect and correct spectral alterations of signals by employing statistical characterization of waveforms and a pattern-recognition procedure using simple Artificial Neural Networks. The proposed strategy implements very-fast routines with a computational cost proportional to the number of signal samples, being convenient for applications in embedded environments with limited computational capabilities or fast real-time control tasks. We use the proposed algorithms to correct spectral amplitude attenuations in a pulse-like waveform with a sinc profile as an application example.
\end{abstract}

\begin{IEEEkeywords}
signal processing, artificial neural networks, statistical characterization, real-time systems
\end{IEEEkeywords}

\section{Introduction}\label{sec:intro}

A spectral characterization by Fourier transform is a ubiquitous and powerful tool used to extract behaviors and features of waveforms \cite{NghiaDo98,Hossen13_JCMSE,Hui14,Lin18}, particularly of the response of signals affected by physic systems, as it is well known. The response spectrum gives accurate information about how each frequency component of the signal is altered by the system. The computational cost of the Fourier transform is proportional to $N_{\!s} \log_2(N_{\!s})$ (for FFT) \cite{NumRec_3ed_C12-C13}, where $N_{\!s}$ is the number of signal samples.
Nevertheless, in some cases where it is necessary to extract general properties of the signals, the Fourier analysis could have a high computational cost compared with other techniques applied directly to the time domain. In particular, due to this cost, a complete characterization using frequency-domain methods could be inefficient in an embedded environment with limited computational capability or when a very-fast procedure is required.
Accordingly, the general properties of the signals can be measured using statistical methods \cite{Book_Hirsch92,Hirsch92,Pakhomov03,Hossen06,Hossen07,Esmael12,SegForFarAn13,Hossen14,Oshea18,Zhang20}, where moments and cumulants are calculated with a computational cost proportional to $N_{\!s}$.\\

In this work, we use statistical techniques to characterize alterations or deformations in the signals and their spectra due, for instance, to non-idealities in the communication channels or amplifiers. We aim to identify such alterations and compensate them by applying modifications in the input signal; then, the response or system output signal would be the desired original signal without deformations.
We want to apply these proceedings in a real-time control strategy, which can be supported in an embedded processor; thus, for this purpose, we choose a time-domain statistical method as a fast feature extraction procedure. In particular, we use the Extended SSC technique (ESSC) developed by the authors \cite{BusSeg2022} as an extension of the SSC (Statistical Signal Characterization) method created by H.L. Hirsch \cite{Book_Hirsch92,Hirsch92} to characterize the signal. Such an extended method uses the 4-parameters strategy proposed by Hirsch and adds some moments and cumulants of the signal, derivative and integral, conforming to a 30-parameters characterization procedure.
The signals we want to describe have pulse-like forms; they are used in applications such as NMR (Nuclear Magnetic Resonance) or MRI (Magnetic Resonance Imaging), radar or communication techniques, etc. In such applications, shaped-pulses methods accomplish different actions, and the efficiency of these methods depends on the precision of the signal form.
Thus, the correction strategy is performed by recognizing which waveform is more similar to the response signal between a set of waveforms with known spectral alterations; then, we can modify the input and compensate for such an alteration. This recognition task can be achieved using simple Artificial Neural Networks (ANNs), as we will describe in the following sections.\\

The remainder of this paper is organized as follows: \sect{sec:td-met} introduces the method used in this work to characterize spectral alterations by employing a time-domain procedure. \sect{sec:app_sac} presents an application of such a method proposing an algorithm to correct spectral amplitude attenuations affecting a signal or waveform. Accordingly, \sect{sec:met_sac} explains the spectral-correction strategy in detail, which implements statistical characterization of the altered waveforms and a pattern-recognition procedure using ANNs; then, \sect{sec:ex_sac_sinc} applies the concepts to correct alterations in a ${\rm sinc}$-pulse waveform as an example. Finally, \sect{sec:disc_conc} discusses the results and presents the conclusions.

\section{Time-domain method to characterize spectral bands of a signal}\label{sec:td-met}

The strategy chosen to detect alterations in the signal spectrum is to split such a spectrum into bands; then, we inspect each frequency band looking for deviations or deformations by comparing it with an unaltered or non-deformed spectrum or signal. This kind of strategy, splitting the spectrum into bands and observing the resulted signals on the time domain, is similar to obtaining a wavelet development for the signal and is common practice in the literature \cite{Hossen13_6ICBEI,Hossen13_THC}.
We want to correct possible amplitude spectrum deformations of the signals; thus, we present the method and results showing amplitude-frequency graphs, but the concepts can also apply to the phase spectrum. Accordingly, in the case of phase corrections, we can use a similar technique that in the amplitude case (or for real and imaginary spectrum parts).\\
Therefore, we can develop a signal $f(t)$ using filters to extract each frequency band, this is
\begin{equation}\label{eq:DesEsp}
\begin{split}
f(t) &=  \opc{F}^{-1}_{\nu \rightarrow t}\left\{{\rm F}(\nu)\,\sum_n \fnFC{n}(\nu)\right\}\\
&= \sum_n \opc{F}^{-1}_{\nu \rightarrow t}\left\{{\rm F}(\nu)\,\fnFC{n}(\nu)\right\}
= \sum_n g^{(n)}_C(t),
\end{split}
\end{equation}
with
\begin{equation}\label{eq:SumaFa1}
\sum_n \fnFC{n}(\nu) = 1,
\end{equation}
and
\begin{equation}\label{eq:Def_gnC}
g^{(n)}_C(t) \equiv \opc{F}^{-1}_{\nu \rightarrow t}\left\{{\rm F}(\nu)\,\fnFC{n}(\nu)\right\},
\end{equation}
where ${\rm F}(\nu) \equiv \opc{F}_{t \rightarrow \nu}\left\{f(t)\right\}$ is the Fourier transform of $f(t)$ from the variable $t$ to $\nu \equiv \omega/(2\pi)$, and $\opc{F}^{-1}_{\nu \rightarrow t}\left\{\cdot\right\}$ represents the inverse transform.
In \eq{eq:DesEsp}, the functions $\fnFC{n}(\nu)$ are the spectral behaviors of the filters used to decompose the signal spectrum, and $g^{(n)}_C(t)$ are the functions extracted by such filters from the original signal.
Such spectral decomposition is shown in \fig{fig:DE_Sinc} for a ${\rm sinc}$-pulse waveform (a.1) with its amplitude spectrum (a.2).
In the graphs, we use a normalized time variable ($t$) ranging between 0 and 1; thus, the frequency ($\nu$) scale is reciprocal and arbitrary (e.g., if we have $\un{ms}$ in $t$, then we obtain $\un{kHz}$ in $\nu$, and so on).
We can observe the spectral splitting process in different bands by applying the filters with spectrum $\fnFC{n}(\nu)$ (dotted light green lines in (b.2), (c.2), (d.2), (e.2), and (f.2)), where we obtain the wavelets $g^{(n)}_C(t)$ (solid blue lines in (b.1,b.2), (c.1,c.2), (d.1,d.2), (e.1,e.2), and (f.1,f.2)) by means of the inverse Fourier transform over the spectra ${\rm F}(\nu)\,\fnFC{n}(\nu)$ (\eq{eq:Def_gnC}). Such wavelets develop $f(t)$ (\eq{eq:DesEsp}), as is represented in (g.1).\\

%{8.5}{16.82} {13.85} {6.5}{12.86}
\insertfig{htbp}{fig:DE_Sinc}{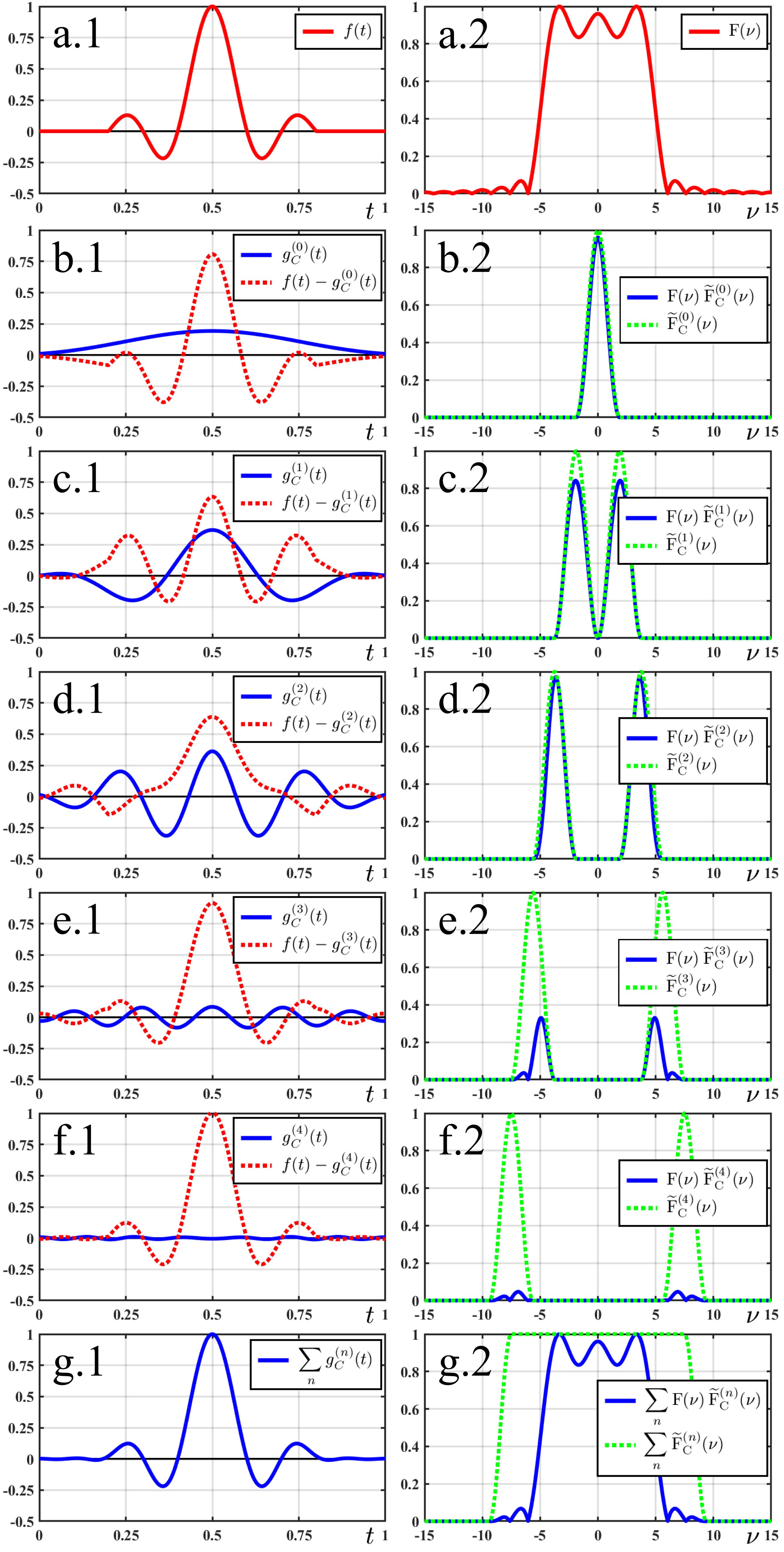}{8.5}{16.82}{Spectral decomposition of a signal (${\rm sinc}$-pulse waveform) into frequency bands, using five filters centered in the range $0 \leq \nu \leq 7.5$. (a.1): original signal $f(t)$, (a.2): ${\rm F}(\nu)$ is the normalized amplitude spectrum of $f(t)$. (b.2, c.2, d.2, e.2, and f.2): $\cos$-profile filters $\fnFC{n}(\nu)$ ($n = \{0,1,2,3,4\}$, dotted light green lines), signal spectra for each band ${\rm F}(\nu)\,\fnFC{n}(\nu)$ (solid blue lines). (g.2): sum of the filter profiles (dotted light green line), sum of the band signal spectra (solid blue line). (b.1, c.1, d.1, e.1, and f.1): decomposition wavelets $g^{(n)}_C(t)$ (inverse Fourier transforms of ${\rm F}(\nu)\,\fnFC{n}(\nu)$, solid blue line), wavelet extraction from the original signal (dotted red lines). (g.1): wavelet sum.}
%{Spectral decomposition of a signal (${\rm sinc}$-pulse waveform) into frequency bands, using five filters centered in the range $0 \leq \nu \leq 7.5$. It is shown the original signal $f(t)$ (a.1) and its normalized amplitude spectrum ${\rm F}(\nu)$ (a.2). The bands are extracted by the $\cos$-profile filter $\fnFC{n}(\nu)$ (with $n = {0,1,2,3,4}$), whose normalized amplitude spectra are shown in dotted light green lines (b.2, c.2, d.2, e.2, and f.2); the sum of all these spectra $\left(\sum_n\fnFC{n}(\nu)\right)$ is equal to 1 in the frequency region to be developed, as it is shown in (g.2) in a dotted light green line. The filter amplitude spectra are obtained by multiplying the spectra $\fnFC{n}(\nu)$ times the signal spectrum ${\rm F}(\nu)$; their normalized amplitude spectra and result waveforms $g^{(n)}_C(t)$ (like wavelets) are shown in solid blue lines in (b.2, c.2, d.2, e.2, and  f.2) and (b.1, c.1, d.1, e.1, and f.1), respectively. The sum of all the extracted spectra $\left(\sum_n{\rm F}(\nu)\,\fnFC{n}(\nu)\right)$ and waveforms $\left(\sum_n g^{(n)}_C(t)\right)$ are shown in solid blue lines in (g.2) and (g.1), respectively, where (g.2) is the normalized amplitude spectrum of (g.1). In dotted red lines in (b.1, c.1, d.1, e.1, and f.1) are shown the result waveforms by resting the original signal minus the wavelets $\left(f(t) - g^{(n)}_C(t)\right)$; these conform to the aw-set with a wavelet attenuation of 100\%.}

The sum of the spectra of the filters $\fnFC{n}(\nu)$ must be 1 (see \eq{eq:SumaFa1}) to develop the signal adequately, with non-introducing changes in it. However, if the amplitude signal spectrum is bounded in frequencies; this is, it is null or with negligible values for frequencies higher than a frequency $\nu_{\!M}$; then, we can approximate the development \numeq{eq:DesEsp} by a finite number $n_{\!M}$ of filters that extract the exact spectrum in the range $\abs{\nu} < \nu_{\!M}$. Such a bounded development is represented by \fig{fig:DE_Sinc}, wherein (g.2) is shown the sum of the filter spectra (dotted light green line) and the frequency region that accomplishes \numeq{eq:SumaFa1}; besides, it is shown the total filtered signal spectrum
$\sum_n{\rm F}(\nu)\,\fnFC{n}(\nu)$ (solid blue line), whose inverse Fourier transform is shown in (g.1).
There is no other theoretical restriction for the spectral profile of $\fnFC{n}(\nu)$.
%, but we want that $g^{(n)}_C(t)$ to be bounded in time (with values close to 0 when $t \rightarrow 0$ or $t \rightarrow 1$). Accordingly, we used profiles without abrupt changes.
Therefore, we chose for $\fnFC{n}(\nu)$ the following $\cos$-profile form
\begin{equation}\label{eq:Def_FCn}
\fnFC{n}(\nu) \!=\! \left\{
\begin{array}{l}
    \!\!\!\frac{1}{2}\left[1+\cos\left(2\pi\nu/\Delta_\nu\right)\right]G_{\Delta_\nu}(\nu), \,n=0,\\
    \!\!\!\frac{1}{2}\left[1+\cos\left(2\pi\nu/\Delta_\nu\right)\right]\\
        \times\left[G_{\Delta_\nu}(\nu\!-\!\nu_{\!n}) \!+\! G_{\Delta_\nu}(\nu\!+\!\nu_{\!n})\right], \,n\,\text{even},\\
    \!\!\!\frac{1}{2}\left[1-\cos\left(2\pi\nu/\Delta_\nu\right)\right]\\
        \times\left[G_{\Delta_\nu}(\nu\!-\!\nu_{\!n}) \!+\! G_{\Delta_\nu}(\nu\!+\!\nu_{\!n})\right], \,n\,\text{odd},
\end{array}
\right.
\end{equation}
with the rectangular function
\begin{equation*}\label{eq:Def_Gn}
G_{\Delta_\nu}(\nu) = \left\{
\begin{array}{l}
    \!\! 1, \;\;\abs{\nu} \leq \Delta_\nu/2.\\
    \!\! 0, \;\;\abs{\nu} > \Delta_\nu/2.
\end{array}
\right.
\end{equation*}
In \eq{eq:Def_FCn}, $\Delta_\nu$ is the period or maximum width of the filters, and $\nu_{\!n}$ is the central frequency of the $n$th filter.
In \fig{fig:DE_Sinc}, we used the filter profile \numeq{eq:Def_FCn}.
The filters must be equispaced with their central frequencies in the correct position to accomplish \eq{eq:SumaFa1}; thus, we have
\begin{equation}\label{eq:Set_Delta_nu_n}
\Delta_\nu = 2\,\nu_{\!M}/n_{\!M}, \quad \nu_{\!n} = n\,\Delta_\nu/2,
\end{equation}
where $\nu_{\!M}$ is the maximum frequency where we want to accomplish strictly \eq{eq:SumaFa1}, and $n_{\!M} > 0$ is the chosen value for the maximum filter number in the spectral development of the signal.\\

Finally, we conclude that we can obtain a time-domain characterization of different spectral bands of a signal by analyzing their associated wavelets $g^{(n)}_C(t)$. In the next section, we use this approach to correct amplitude alterations in the signal spectrum.

\section{Application to spectral amplitude correction of a signal}\label{sec:app_sac}

We aim to correct the spectral deformations of a signal. In order of that, we build a set of waveforms with known alterations (or without any changes) by taking the original signal and applying an amplitude attenuation in a frequency band of the development \numeq{eq:DesEsp}; we obtain
\begin{equation}\label{eq:DesEsp_aten}
\begin{split}
&f_{\!n}(t,\Delta_a) = \opc{F}^{-1}_{\nu \rightarrow t}\bigg\{\!{\rm F}(\nu)\bigg[\sum_{m \ne n} \fnFC{m}(\nu)\\
            &\qquad\qquad\qquad\qquad\qquad\qquad\quad+ (1 \!-\! \Delta_a)\,\fnFC{n}(\nu)\bigg]\!\bigg\}\\
&= \opc{F}^{-1}_{\nu \rightarrow t}\left\{\!{\rm F}(\nu)\left[1 \!-\! \Delta_a\,\fnFC{n}(\nu)\right]\!\right\}
= f(t) \!-\! \Delta_a\,g^{(n)}_C(t),
\end{split}
\end{equation}
where $f_{\!n}(t,\Delta_a)$ is the altered waveform obtained by attenuating the $n$th band of $f(t)$ in a factor $\Delta_a$ ($0 \leq \Delta_a \leq 1$).
We can see from \eq{eq:DesEsp_aten} that such an alteration is equivalent to applying the following band-rejection filter over the signal
\begin{equation}\label{eq:Def_FC}
\fFC{n}(\nu,\Delta_a) \equiv 1 - \Delta_a\,\fnFC{n}(\nu).
\end{equation}
The altered-waveform set $\{f_{\!n}(t,\Delta_a)\}$, with
\[n = \{0,1,2,\dots,n_{\!M}\},\]
and different values of $\Delta_a$, will be called \emph{aw-set}. In \fig{eq:Def_FCn}, we see the set $\left\{f_{\!n}(t,1) = f(t) \!-\! g^{(n)}_C(t)\right\}$
(dotted red lines in b.1, c.1, d.1, e.1, and f.1), where $n = \{0,1,2,3,4\}$; which corresponds to the aw-set with 100\% ($\Delta_a = 1$)
of attenuation in the bands.\\

Therefore, we might identify the type of amplitude spectrum alteration by comparing the aw-set and the system response signal in the time domain; then, we would obtain the $\Delta_a$ factors to be correct in the bands. In the following sections, we propose a correction method using these concepts.

\subsection{Description of the spectral-correction method}\label{sec:met_sac}

The spectral-correction method that we propose consists of identifying the type of amplitude spectral alteration by using an Artificial Neural Network (ANN) to extract the probability of similarity between the system response signal and the members of the aw-set in the time domain. Then, because we know the spectral bands and values of the attenuations of the aw-set, we can obtain a series of correction factors employing such probabilities.
As was commented, the aw-set is obtained by applying over the original signal (without alterations) the band-rejection filters of \eq{eq:Def_FC} in the range $\abs{\nu} < \nu_{\!M}$ and for different attenuations ($\Delta_a$), where each band is configured with a width and central frequency as in \eq{eq:Set_Delta_nu_n}. Then, we train and test the ANN using the algorithm represented in \fig{fig:alg_ANN}.\\

%{17}{7.84}  {13}{6} \insertfigdc{!t}{8.5}{3.92}
\insertfigdc{htbp}{fig:alg_ANN}{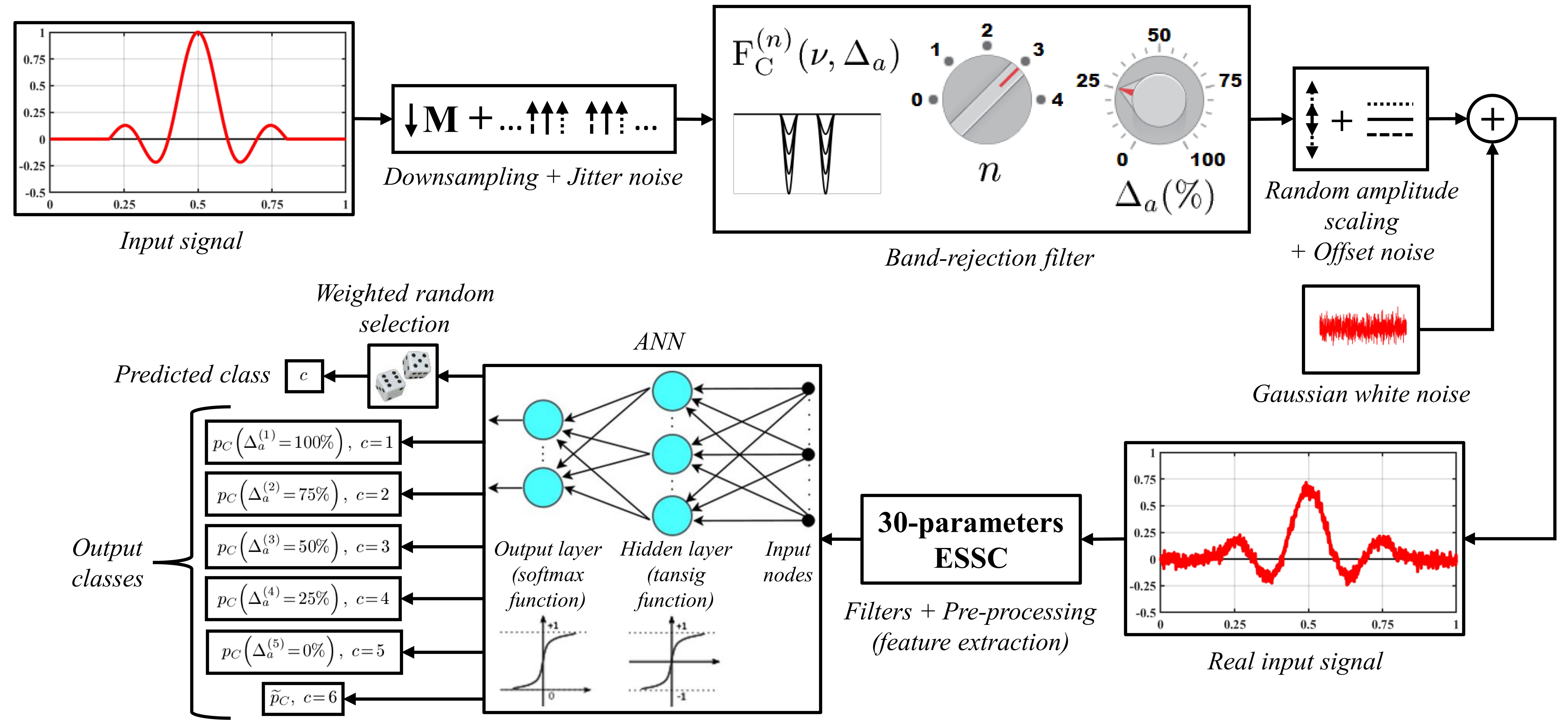}{17}{7.84}{Time-domain algorithm used to train and test the Artificial Neural Networks for the spectral-correction method.}

In \fig{fig:alg_ANN}, the schematic diagram describes the following processes.
First, we generate an ideal pulse \emph{input signal} with a time resolution of 10000 points. Such an ideal signal passes through a \emph{downsampling} stage where the samples are reduced by a factor of ${\rm M}=10$ (decimation) and is applied a uniform-random time \emph{jitter noise} (between 0 and 9 samples) aimed at emulating the effects of an acquisition process. After that, we apply a filter $\fFC{n}(\nu,\Delta_a)$ (\eq{eq:Def_FC}) where we choose the band $n$ and the attenuation factor $\Delta_a$ to obtain a waveform member of the aw-set.
Following the filter stage, we perform a uniform-\emph{random amplitude scaling} (up to 75\% of the maximum value), and an \emph{offset noise} introduces a uniform-random continuous level (up to 5\% of the maximum amplitude value). Then, the \emph{real input signal} is finally obtained by adding a \emph{Gaussian white noise} (with zero mean and the standard deviation set as 5\% of the maximum signal amplitude). Therefore, we have a simulation of a real acquired signal for the altered waveform (band $n$ and attenuation $\Delta_a$) to feed the pre-processing algorithm of the ANN.

The \emph{pre-processing} stage is an algorithm that aims to extract several relevant features of the input signal. Such a stage is a crucial part of the ANN input, where a waveform with $N_{\!s}$ samples is represented by several characteristic features or parameters, reducing the complexity of the processing of the ANN. We use a very-fast statistical method as a pre-processing stage; that is called ESSC (Extended Statistical Signal Characterization) by the authors \cite{BusSeg2022} and is proposed as an extension of the SSC one developed by H.L. Hirsch \cite{Book_Hirsch92,Hirsch92}. This method consists of the calculation of the SSC parameters (4 for each waveform) plus several moments and cumulants of the signal, its derivative and integral, in a total of 30 parameters conforming a `fingerprint' of the waveform. Before the pre-processing stage, the input signal is affected by a median and a mean filter, an offset correction, and a pulse detection routine. Besides, the signal is normalized in amplitude and time, and it is adequate for entry in the ESSC algorithm. That \emph{filters+pre-processing} block is called \textbf{30-parameters ESSC} in \fig{fig:alg_ANN}, and its computational cost is proportional to $N_{\!s}$.

We implement a feed-forward back-propagation ANN to solve the pattern-recognition task between the different waveform members of the aw-set.
The ANN structure (see the \emph{ANN} block in \fig{fig:alg_ANN}) is formed by an input layer with 30 \emph{input nodes}, where this amount corresponds to the ESSC parameter entries. Then, a \emph{hidden layer} has between 5 and 40 neurons with a \emph{tansig} activation function.
At the end of the network, an \emph{output layer} includes neurons and terminals as the number of \emph{output classes} to be predicted ($n_c$). Such output neurons use the \emph{softmax} activation function and implement the \emph{one-hot encoding method}. The softmax's output is an $n_c$-elements array with the prediction level for each class; this is a range of values between 0 and 1. Such a level is the probability of association of the input parameters (or waveform) with such a class; thus, we use a \emph{weighted random selection} employing these probabilities to choose the \emph{predicted class} ($c\,$) for that input (i.e., the output result of the ANN).
To define the number of neurons in the hidden layer, we trained the network with a different quantity of hidden neurons and we evaluated the average \emph{cross-entropy} of the softmax's output; then, we used the ANN configuration and the number of hidden neurons with the lowest cross-entropy value.\\

The process depicted in \fig{fig:alg_ANN} is used to train an ANN per frequency band (i.e., for a filter $\fFC{n}(\nu,\Delta_a)$ with a fixed value of $n$).
We note that an equal \emph{input signal} with the same filter choice produces different \emph{real input signals} in the \emph{pre-processing} stage due to the randomness introduced by several noise sources. Therefore, the training of the ANN is performed by exhaustively running the algorithm shown in \fig{fig:alg_ANN}.\\
We used the following five attenuation values
\begin{equation}\label{eq:Val_Delta_a}
\begin{split}
\Delta_a = \left\{\Delta^{(c)}_a\right\} &= \left\{1\,(100\%),\, 0.75\,(75\%), \right.\\
&\qquad \left. 0.5\,(50\%),\, 0.25\,(25\%),\, 0\,(0\%)\right\}
\end{split}
\end{equation}
to obtain the aw-set of each band:
\begin{equation}\label{eq:aw-set}
\left\{f_{\!n}\left(t,\Delta^{(c)}_a\right) = f(t) \!-\! \Delta^{(c)}_a\,g^{(n)}_C(t)\right\},
\end{equation}
with $c = \{1,2,3,4,5\}$. The 5th class is the non-attenuated waveform, $\Delta^{(5)}_a = 0\,(0\%)$.
A network is trained using a dataset with arrays of 30-parameters (ESSC) as inputs. Such a dataset corresponds to the parameters of all the waveforms in the aw-set \numeq{eq:aw-set}, and  it is provided the class to which each waveform belongs or \emph{true class}. Besides, we used a 6th class to identify that the alteration is in another band. In order of that, we train the same network using a dataset with altered waveforms from other bands or filters, where $\Delta_a = \{1,\, 0.75,\, 0.5,\, 0.25\}$ ($\Delta_a = 0$ is yet included in the 5th class). Accordingly, the outputs of the softmax function are the probabilities $p_C\!\left(\Delta^{(c)}_a\right)$, for $c=1$ to $5$, and $\widetilde{p}_C$ (the attenuation is not in this band), which is defined for $c=6$. Then
\[\sum_{c=1}^{5} p_C\!\left(\Delta^{(c)}_a\right) + \widetilde{p}_C = 1, \;\;\text{and}\;\; n_c = 6.\]

The performance of the ANN to recognize different waveforms of the aw-set \numeq{eq:aw-set} can be tested using a dataset similar to the training one and comparing the \emph{true class} of a waveform with the output \emph{predicted class} of the network. Such a comparison is commonly shown in the form of a \emph{confusion matrix}, where the rows (columns) represent the \emph{true classes} and the columns (rows) represent the \emph{predicted classes}; then, the matrix element values are the number of times that a \emph{predicted class} is obtained as an output of the ANN for a fixed \emph{true class}.
We could obtain a different \emph{predicted class} value for the same waveform under different algorithm runs; due to the \emph{weighted random selection} process. Therefore, we calculate an average \emph{confusion matrix} from several runs with the same waveform to represent the network performance adequately.\\

After the building, training, and testing of the ANNs for all the bands, we may use the networks to implement the correction algorithm to the input (or original) signal. Such an algorithm is described by the schematic diagram in \fig{fig:alg_Correct}.
The \emph{real input signal} is the original waveform affected by an arbitrary spectral amplitude attenuation (or alteration) due to non-idealities, for instance, in the transmission physic system. In our approach, we suppose that general spectral attenuations due to the system can be expressed or very well approximated by
\begin{equation}\label{eq:FDef_sys}
{\rm F_S}(\nu) = \prod_n \left(1 - \overline{\Delta}_{a,n}\,\fnFC{n}(\nu)\right);
\end{equation}
this is, the system spectral response can be developed by a product of band-rejection filters of type \numeq{eq:Def_FC}.\\
The \textbf{30-parameters ESSC} block extracts the waveform features using very-fast filters and time-domain statistical pre-processing of the signal; then, a 30-parameters array is sent to each ANN to be analyzed.
We have as many ANNs as the number of bands that we want to split the frequency range of the signal, where the \emph{ANNn} ($n = 0,1,\dots$) is trained using the aw-set obtained by applying a band-rejection filter $\fFC{n}(\nu,\Delta_a)$ with a width and central frequency defined in \eq{eq:Set_Delta_nu_n} and with the different attenuations of \eq{eq:Val_Delta_a}.\\
We obtain a \emph{mean correction factor} for the band $n$ using the probability of each class of the output softmax function of \emph{ANNn}, this is
\begin{equation}\label{eq:Prom_Delta_an}
\overline{\Delta}_{a,n} = \sum^{n_c}_{c=1} \Delta^{(c)}_a\;p_C^{(n)}\!\!\left(\Delta^{(c)}_a\right)
= \sum^{4}_{c=1} \Delta^{(c)}_a\;p_C^{(n)}\!\!\left(\Delta^{(c)}_a\right),
\end{equation}
with $n_c = 6$, where for simplicity we define $p_C^{(n)}\!\!\left(\Delta^{(6)}_a\right) \equiv \widetilde{p}_C$, and
$\Delta^{(6)}_a = \Delta^{(5)}_a = 0$. In \eq{eq:Prom_Delta_an}, $p_C^{(n)}$ represents the probability associated with the network or band $n$.\\
Finally, we use the correction factors \numeq{eq:Prom_Delta_an} to obtain the \emph{corrected signal}
\begin{equation}\label{eq:CorrSig}
\begin{split}
\widehat{f}(t) &= \opc{F}^{-1}_{\nu \rightarrow t}\left\{{\rm F}(\nu)\,\prod_n \left(1 - \overline{\Delta}_{a,n}\,\fnFC{n}(\nu)\right)^{-1}\right\}\\
&= \opc{F}^{-1}_{\nu \rightarrow t}\left\{{\rm F}(\nu)\,\prod_n \sum_{q_{n}=\,0}^{\infty}\overline{\Delta}^{\,q_{n}}_{a,n}\,\left(\fnFC{n}(\nu)\right)^{q_{n}}\right\}\\
&= \sum_{n} \left(1+\overline{\Delta}_{a,n}\right) g^{(n)}_C(t)\\
&\qquad\qquad + \sum_{q_0,\,q_1,\cdots}^{*} \left(\prod_n \overline{\Delta}^{\,q_n}_{a,n}\right)\,g^{\{q_0,\,q_1,\cdots\}}_C(t),
\end{split}
\end{equation}
where are defined the following wavelets
\begin{equation}\label{eq:Def_gqnC}
g^{\{q_0,\,q_1,\cdots\}}_C(t) \equiv \opc{F}^{-1}_{\nu \rightarrow t}\left\{{\rm F}(\nu)\,\prod_n \left(\fnFC{n}(\nu)\right)^{q_n}\right\},
\end{equation}
and $\sum_{q_0,\,q_1,\cdots}^{*}$ represents a sum excluding the indices
\begin{equation*}
\begin{split}
&\{q_0,\,q_1, \cdots, q_n, \cdots\} = \big\{\{0,0,\cdots,0,\cdots\};\\
&\qquad\quad\{1,0,\cdots,0,\cdots\}; \cdots; \{0,0,\cdots,1,\cdots\}; \cdots\big\}.
\end{split}
\end{equation*}
The spectrum of the corrected waveform \numeq{eq:CorrSig} is defined to cancel the system spectral response \numeq{eq:FDef_sys}; thus, we obtain the spectrum ${\rm F}(\nu)$ of the original waveform by multiplying the Fourier transform of the corrected signal \numeq{eq:CorrSig} times the system spectral function \numeq{eq:FDef_sys}.
Therefore, the arbitrary signal spectral attenuation is corrected by using the time-domain development \numeq{eq:CorrSig} as the input signal,
where the compensation factors regulate the amplitude of the wavelet components \numeq{eq:Def_gnC} and \numeq{eq:Def_gqnC}.
Such a development can be approximated by keeping the terms with lower power of $\overline{\Delta}_{a,n}$, whose accuracy will depend on the values of these factors.
Consequently, if we introduce the waveform \numeq{eq:CorrSig} into the analyzed physic system, the system output will be the original signal.\\

%{8.5}{13.54}    {7}{11.2}   {6}{9.56}  {4.4}{7}
\insertfig{htbp}{fig:alg_Correct}{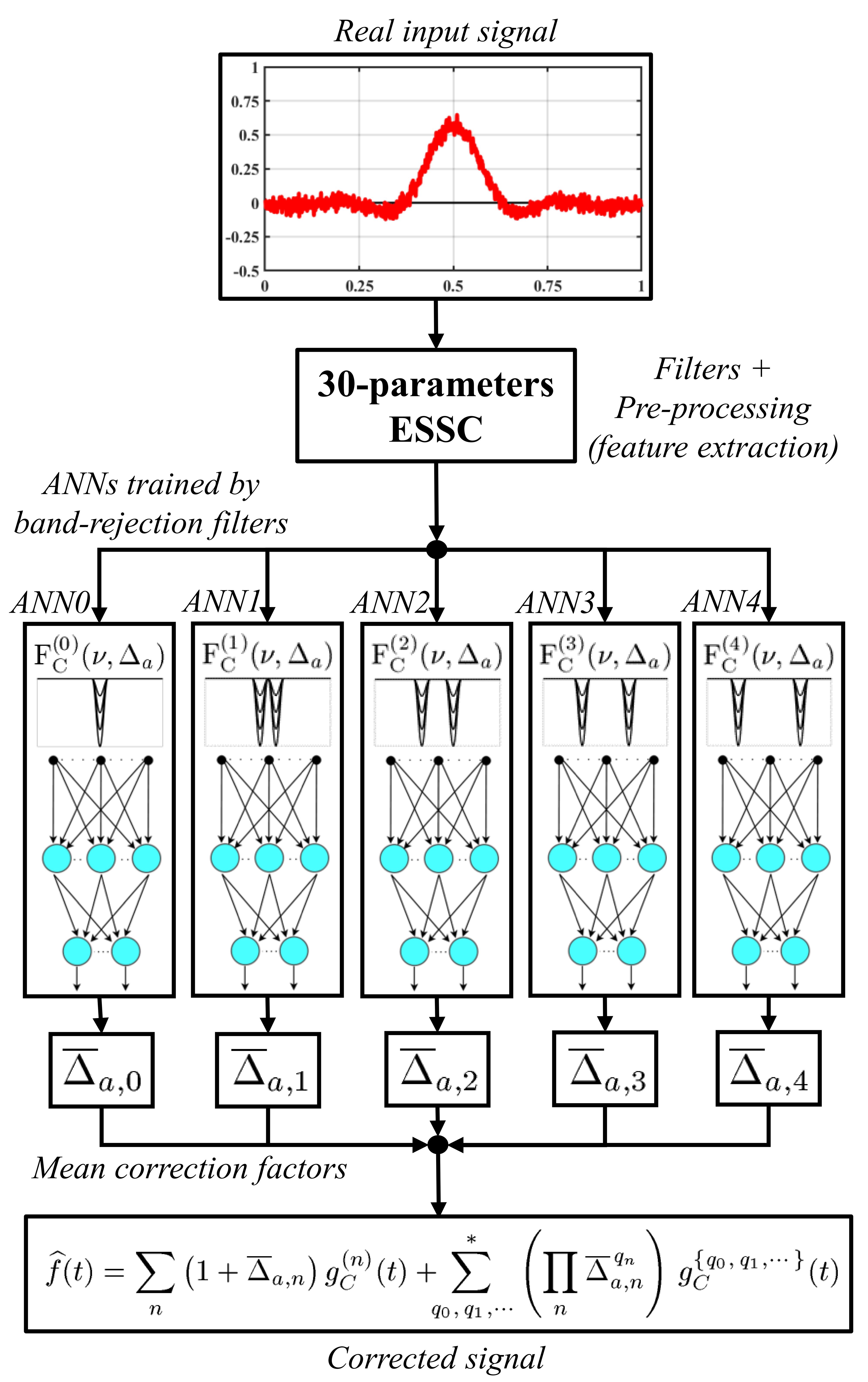}{8.5}{13.54}{Spectral-correction algorithm using time-domain recognition of the altered-wavelet set by employing ANNs.}

In summarizing, the proposed spectral amplitude correction method consists of the following steps
\renewcommand{\theenumi}{\alph{enumi}}
\begin{enumerate}
  \item Define the maximum frequency $\nu_{\!M}$ and the number of filters or bands $n_{\!M}$ to resolve the signal spectrum.
  \item Train and test the ANNs generating the aw-set \numeq{eq:aw-set} of each band by the algorithm of \fig{fig:alg_ANN}.
  \item Apply the algorithm of \fig{fig:alg_Correct} to an arbitrary altered input signal calculating the \emph{mean correction factor} of each frequency band (\eq{eq:Prom_Delta_an}).
  \item Compensate the original signal with the correction factors obtaining the \emph{corrected signal} (\eq{eq:CorrSig}).
  \item Use the corrected signal as system input, and the system output would be the original signal.
\end{enumerate}

As an example, we use the developed algorithms to correct spectral attenuations in a ${\rm sinc}$-pulse waveform in the following section.\\

%\[\begin{array}{c}
%    c \\\\\\
%    p_C\!\left(\Delta^{(1)}_a \!=\! 100\%\right),\; c \!=\! 1 \\\\
%    p_C\!\left(\Delta^{(2)}_a \!=\! 75\%\right),\; c \!=\! 2 \\\\
%    p_C\!\left(\Delta^{(3)}_a \!=\! 50\%\right),\; c \!=\! 3 \\\\
%    p_C\!\left(\Delta^{(4)}_a \!=\! 25\%\right),\; c \!=\! 4 \\\\
%    p_C\!\left(\Delta^{(5)}_a \!=\! 0\%\right),\; c \!=\! 5 \\\\
%    \widetilde{p}_C,\; c \!=\! 6
%  \end{array}
%\]
%\[\fFC{0}(\nu,\Delta_a)\;\;\fFC{1}(\nu,\Delta_a)\;\;\fFC{2}(\nu,\Delta_a)\;\;\fFC{3}(\nu,\Delta_a)\;\;\fFC{4}(\nu,\Delta_a)\]
%\[\overline{\Delta}_{a,0} \quad \overline{\Delta}_{a,1} \quad \overline{\Delta}_{a,2} \quad \overline{\Delta}_{a,3} \quad \overline{\Delta}_{a,4}\]
%\[\Delta^{(1)}_a \!=\! 100\% \quad \Delta^{(2)}_a \!=\! 75\% \quad \Delta^{(3)}_a \!=\! 50\% \quad \Delta^{(4)}_a \!=\! 25\%\]
%\[\overline{\Delta}_{a,n} = \sum^{(n)}_c \Delta^{(c)}_a\;p_C\!\left(\Delta^{(c)}_a\right)\]
%\[\widehat{f}(t) = \sum_n (1+\overline{\Delta}_{a,n})\,g^{(n)}_C(t)\]

\subsection{Example for a ${\rm sinc}$-pulse waveform}\label{sec:ex_sac_sinc}

In this section, we used the method proposed in \sect{sec:met_sac} to detect and correct alterations in a ${\rm sinc}$-pulse waveform
(\fig{fig:DE_Sinc} (a.1), and (a.2)), as an application example of the developed algorithm.
We chose to split the signal spectrum into bands as in \fig{fig:DE_Sinc} with that wavelet decomposition $\left\{g^{(n)}_C(t)\right\}$ (b.1, c.1, d.1, e.1, and f.1), where we selected the maximum frequency $\nu_{\!M} = 7.5$, and band number $n_{\!M} = 4$.
Thus, we obtain from \eq{eq:Set_Delta_nu_n}: $\Delta_\nu = 3.75$, $\nu_{\!n} = 1.875 \times n$.\\

The aw-sets $\left\{f_{\!n}\left(t,\Delta^{(c)}_a\right)\right\}$ \numeq{eq:aw-set} obtained for the band numbers $n = \{0,1,2,3,4\}$ are shown in \figs{fig:Sinc_C0}, \numfig{fig:Sinc_C1}, \numfig{fig:Sinc_C2}, \numfig{fig:Sinc_C3}, and \numfig{fig:Sinc_C4}. In those figures, we can observe in (a.1, b.1, c.1, and d.1) the waveforms for the following classes and attenuation factors: $c = \{1,2,3,4\}$, $\left\{\Delta^{(c)}_a\right\} = \{100\%, 75\%, 50\%, 25\%\}$. The original waveform, shown in \fig{fig:DE_Sinc} (a.1), corresponds to $c = 5$ and $\Delta^{(5)}_a = 0\%$, and it is shared between all the bands. We show in solid red lines in (a.2, b.2, c.2, and d.2) the normalized amplitude spectra of the waveforms. Such spectra are obtained by multiplying the original waveform spectrum ${\rm F}(\nu)$, shown in \fig{fig:DE_Sinc} (a.2), times the filter profiles $\fFC{n}\left(\nu,\Delta^{(c)}_a\right)$, shown in dotted light green lines in (a.2, b.2, c.2, and d.2). The waveforms are generated by the inverse Fourier transform of such spectra.\\

%{8.5}{6.67}   {7}{5.5}  {5.1}{4}
\insertfig{htbp}{fig:Sinc_C0}{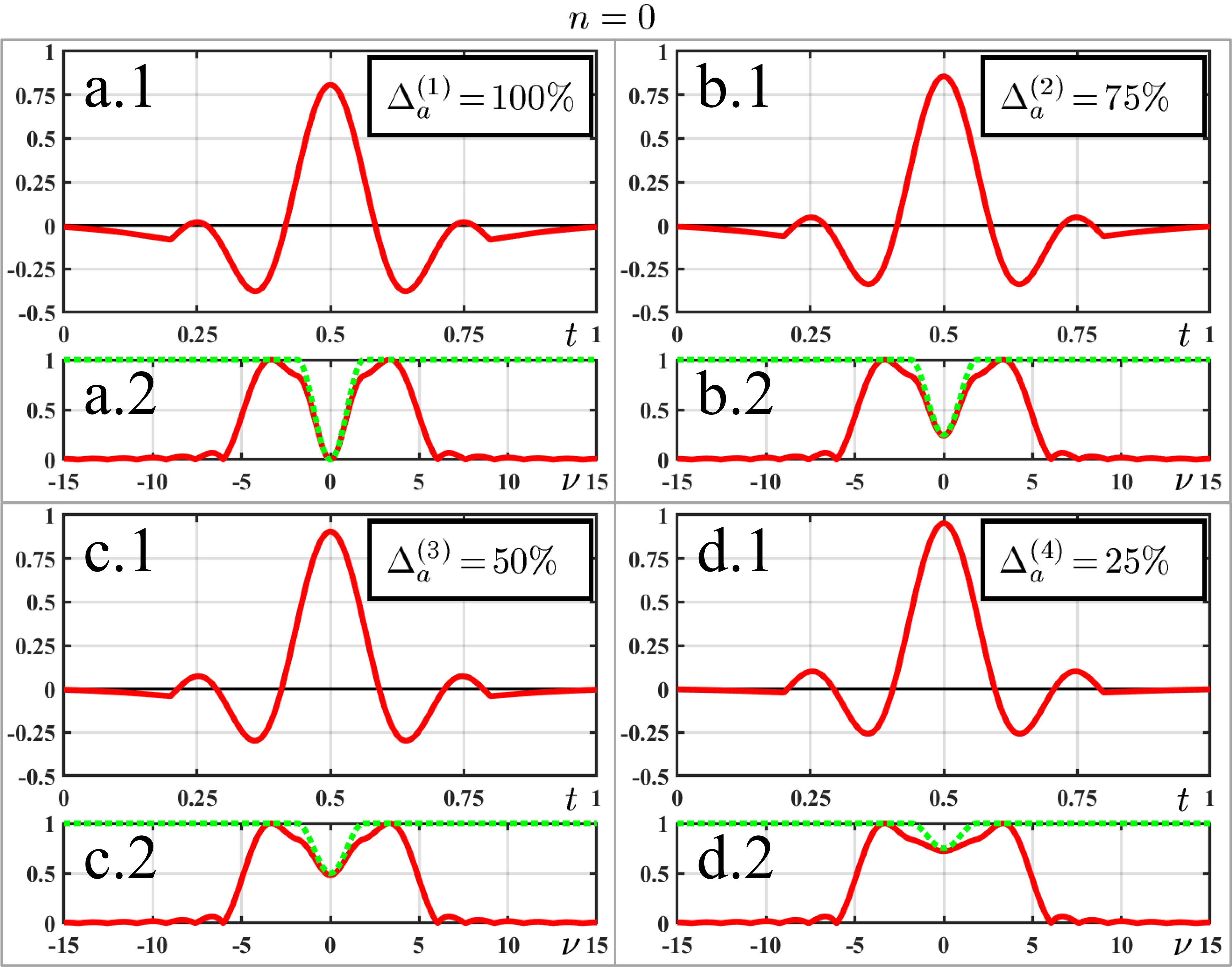}{8.5}{6.67}{Aw-set $\{f_{0}(t,\Delta^{(c)}_a)\}$ ($c = \{1,2,3,4\}$) used in the spectral correction of the band $n = 0$ for a {\rm sinc}-pulse signal. (a.1, b.1, c.1, and d.1): waveforms. (a.2, b.2, c.2, and d.2): normalized amplitude spectra of the waveforms (solid red lines), filter profiles $\fFC{0}(\nu,\Delta^{(c)}_a)$ (dotted light green lines).}
%{Aw-set $\{f_{0}(t,\Delta^{(c)}_a)\}$ (with $c = \{1,2,3,4\}$) used in the spectral correction of the band $n = 0$ for a {\rm sinc}-pulse signal, where in solid red lines are shown the waveforms (a.1, b.1, c.1, and d.1) and their respective normalized spectra (a.2, b.2, c.2, and d.2). The filter profiles $\fFC{0}(\nu,\Delta^{(c)}_a)$ are shown in dotted light green lines (a.2, b.2, c.2, and d.2).}
%We set up the following attenuation factors: $\Delta^{(1)}_a \!=\! 100\%$ (a.1, a.2), $\Delta^{(2)}_a \!=\! 75\%$ (b.1, b.2), $\Delta^{(3)}_a \!=\! 50\%$ (c.1, c.2), and $\Delta^{(4)}_a \!=\! 25\%$ (d.1, d.2).

\insertfig{htbp}{fig:Sinc_C1}{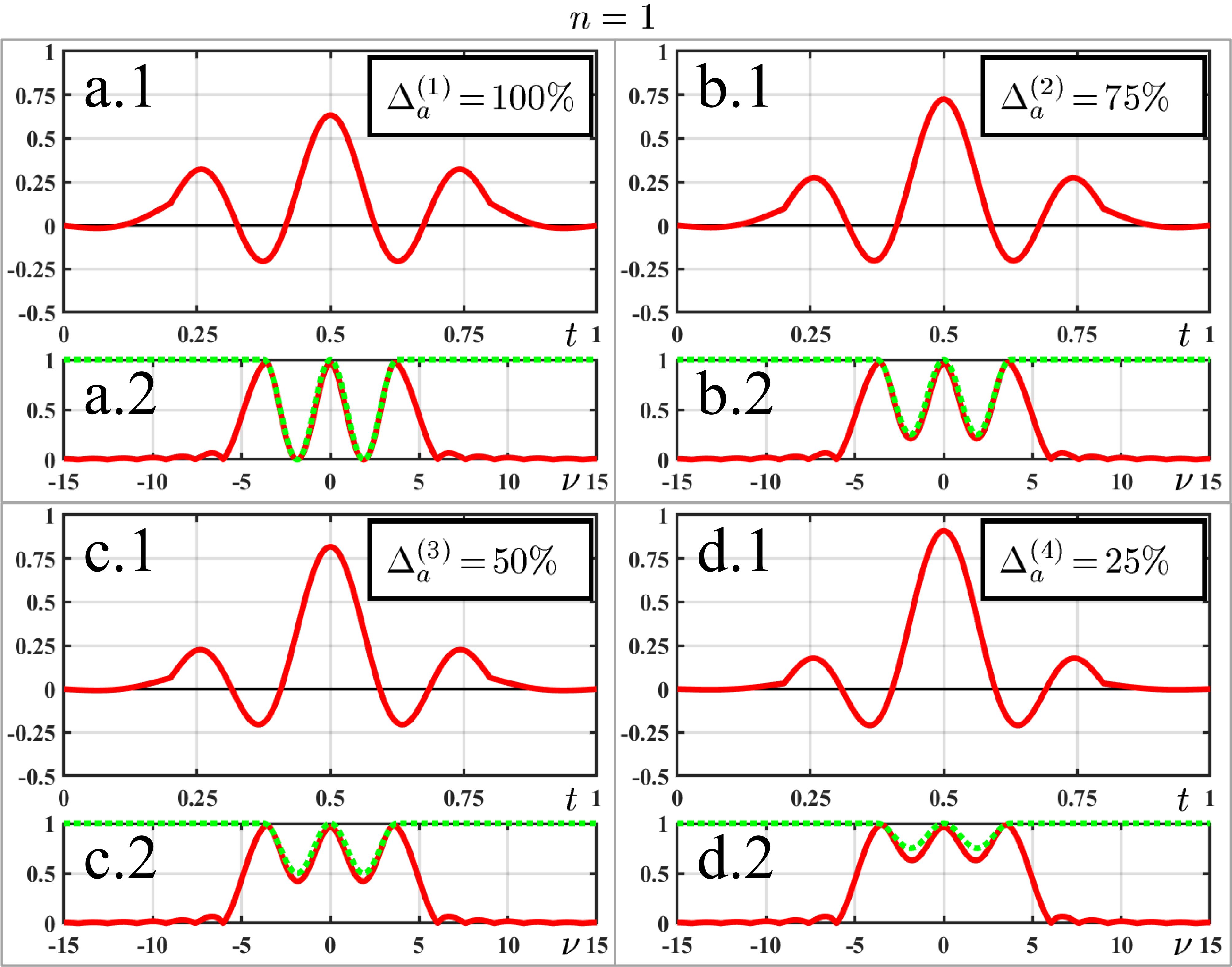}{8.5}{6.67}{Aw-set $\{f_{1}(t,\Delta^{(c)}_a)\}$ for $n = 1$ ($\fFC{1}(\nu,\Delta^{(c)}_a)$). Similar to \fig{fig:Sinc_C0}.}
%{Aw-set $\left\{f_{1}\left(t,\Delta^{(c)}_a\right)\right\}$ (with $c = \{1,2,3,4\}$) used in the spectral correction of the band $n = 1$ for a {\rm sinc}-pulse signal, where in solid red lines are shown the waveforms (a.1, b.1, c.1, and d.1) and their respective normalized spectra (a.2, b.2, c.2, and d.2). The filter profiles $\fFC{1}\left(\nu,\Delta^{(c)}_a\right)$ are shown in dotted light green lines (a.2, b.2, c.2, and d.2).}

\insertfig{htbp}{fig:Sinc_C2}{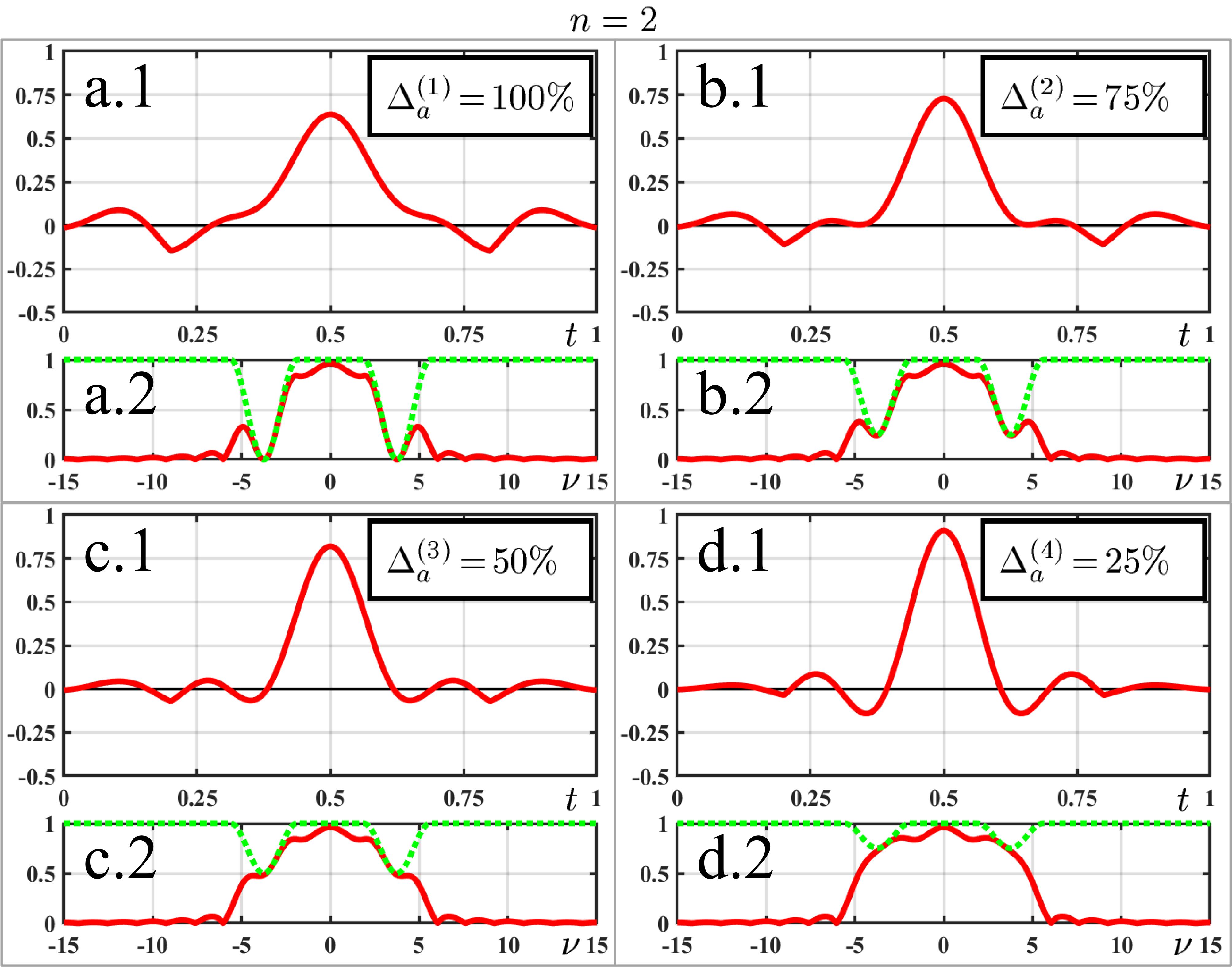}{8.5}{6.67}{Aw-set $\{f_{2}(t,\Delta^{(c)}_a)\}$ for $n = 2$ ($\fFC{2}(\nu,\Delta^{(c)}_a)$). Similar to \fig{fig:Sinc_C0}.}
%{Aw-set $\left\{f_{2}\left(t,\Delta^{(c)}_a\right)\right\}$ (with $c = \{1,2,3,4\}$) used in the spectral correction of the band $n = 2$ for a {\rm sinc}-pulse signal, where in solid red lines are shown the waveforms (a.1, b.1, c.1, and d.1) and their respective normalized spectra (a.2, b.2, c.2, and d.2). The filter profiles $\fFC{2}\left(\nu,\Delta^{(c)}_a\right)$ are shown in dotted light green lines (a.2, b.2, c.2, and d.2).}

\insertfig{htbp}{fig:Sinc_C3}{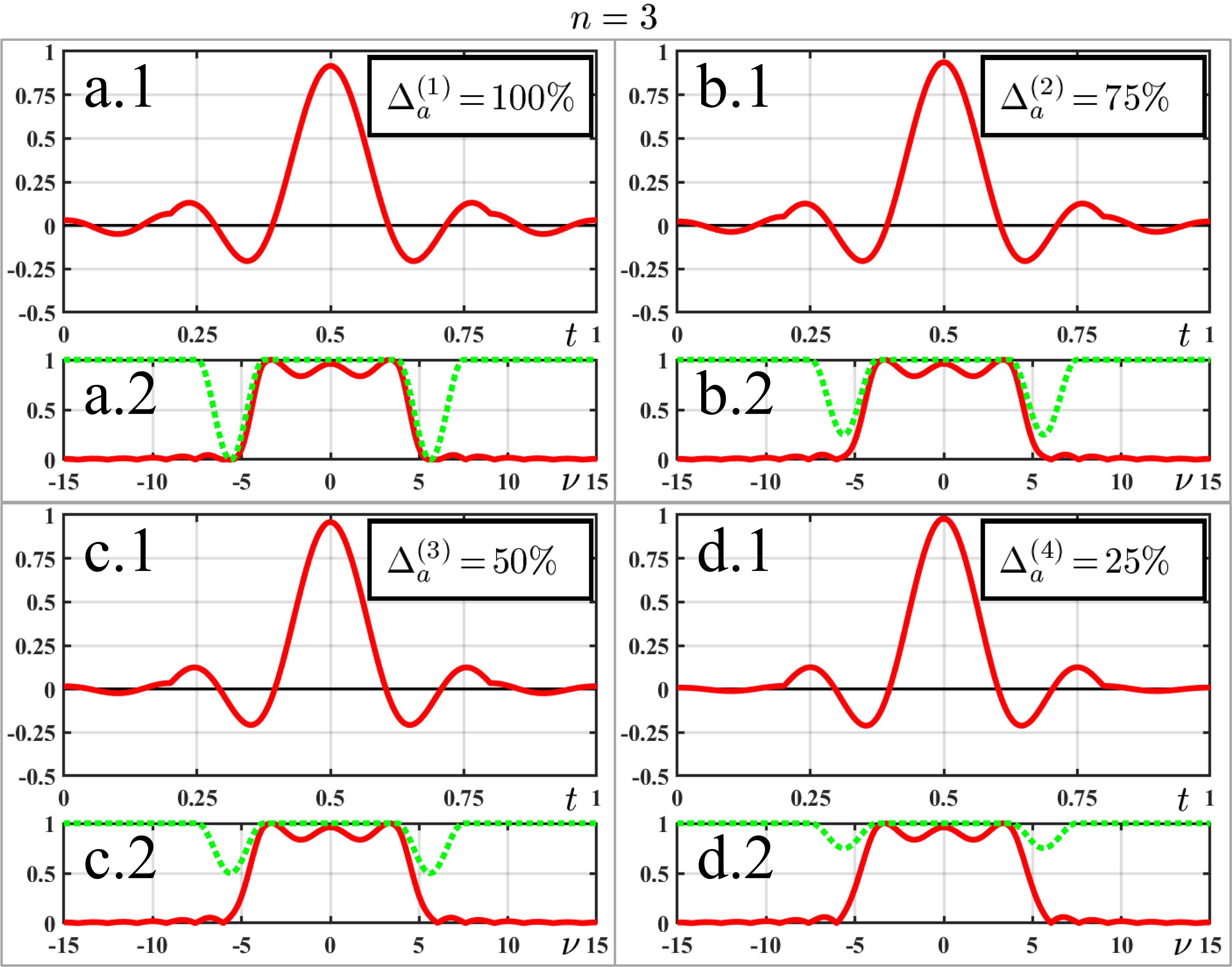}{8.5}{6.67}{Aw-set $\{f_{3}(t,\Delta^{(c)}_a)\}$ for $n = 3$ ($\fFC{3}(\nu,\Delta^{(c)}_a)$). Similar to \fig{fig:Sinc_C0}.}
%{Aw-set $\left\{f_{3}\left(t,\Delta^{(c)}_a\right)\right\}$ (with $c = \{1,2,3,4\}$) used in the spectral correction of the band $n = 3$ for a {\rm sinc}-pulse signal, where in solid red lines are shown the waveforms (a.1, b.1, c.1, and d.1) and their respective normalized spectra (a.2, b.2, c.2, and d.2). The filter profiles $\fFC{3}\left(\nu,\Delta^{(c)}_a\right)$ are shown in dotted light green lines (a.2, b.2, c.2, and d.2).}

\insertfig{htbp}{fig:Sinc_C4}{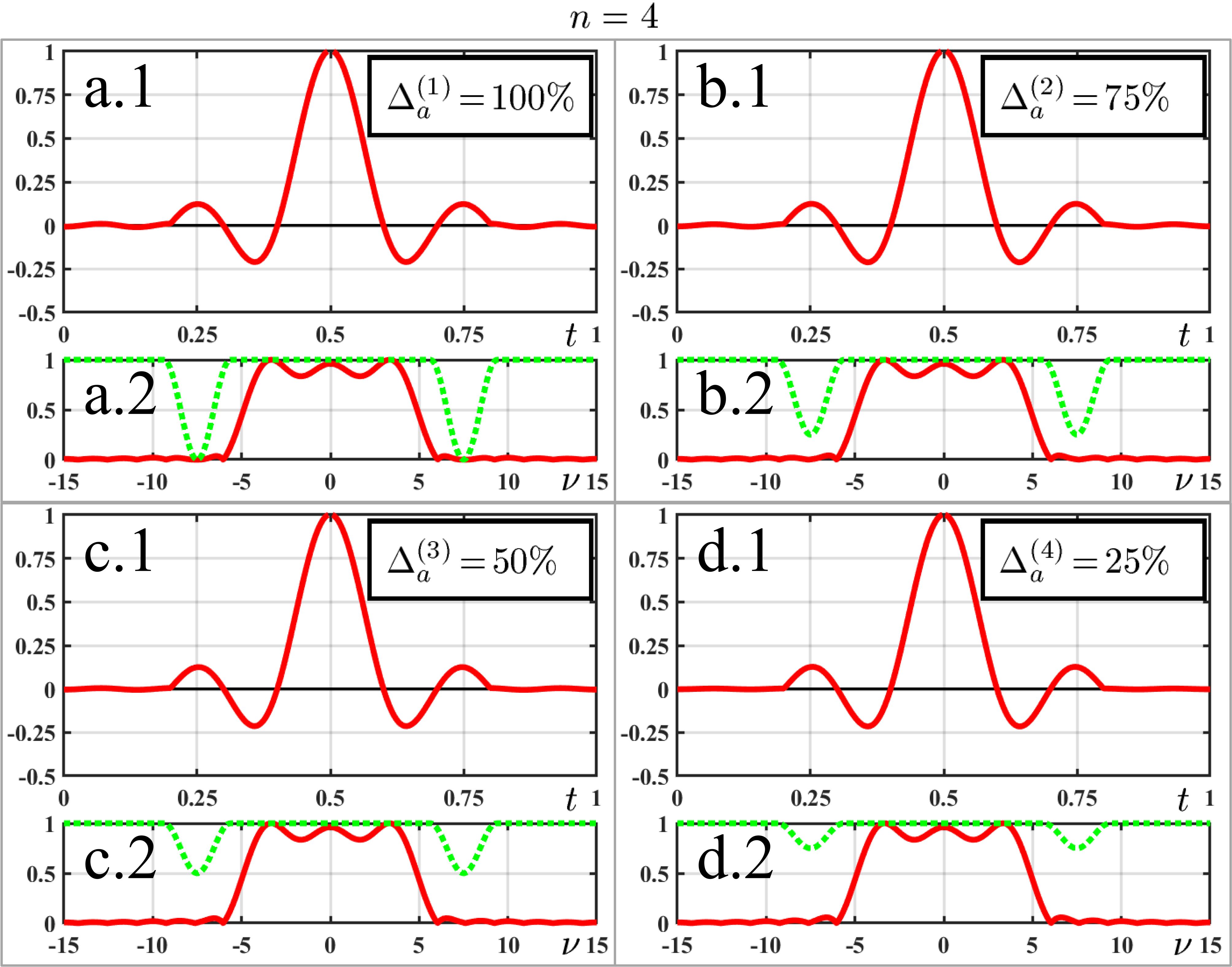}{8.5}{6.67}{Aw-set $\{f_{4}(t,\Delta^{(c)}_a)\}$ for $n = 4$ ($\fFC{4}(\nu,\Delta^{(c)}_a)$). Similar to \fig{fig:Sinc_C0}.}
%{Aw-set $\left\{f_{4}\left(t,\Delta^{(c)}_a\right)\right\}$ (with $c = \{1,2,3,4\}$) used in the spectral correction of the band $n = 4$ for a {\rm sinc}-pulse signal, where in solid red lines are shown the waveforms (a.1, b.1, c.1, and d.1) and their respective normalized spectra (a.2, b.2, c.2, and d.2). The filter profiles $\fFC{4}\left(\nu,\Delta^{(c)}_a\right)$ are shown in dotted light green lines (a.2, b.2, c.2, and d.2).}

We trained and tested the ANN of each band using a dataset of 1000 elements (30-parameters arrays, ESSC) for each waveform belonging to classes 1 to 5, indicating the corresponding class. Accordingly, we run 1000 times the algorithm of \fig{fig:alg_ANN} using the same waveform (\emph{input signal}) from the aw-set to obtain a series of 30-parameters arrays, conforming to the dataset of such a waveform. A class enumerated as $c = 6$ is included identifying that the alteration is in another band. The 6th dataset is formed by taking 250 random elements from the dataset of the remaining bands, for a total of 1000 ones, where each 250-element set has the following distribution: three groups of 62 with 100\%, 75\%, and 50\% of attenuation, a group of 64 with 25\%. Therefore, we have a complete dataset of 6000 elements for all the classes and each band.\\

In \fig{fig:CM}, we show the average confusion matrices representing the performance of the ANNs for the recognition task of the aw-set of each frequency band. The testing procedures of the ANNs were performed using a dataset of 1000 elements per output class, similar to the training one. We averaged the results of 50 matrices to obtain the mean matrices. The off-diagonal elements are averaged, and the diagonal ones are obtained resting 1000 minus the sum of the averaged off-diagonal in the same row. The labels indicate the attenuation factors $\Delta^{(c)}_a$ for the classes $c = \{1,2,3,4,5\}$, and {\rm NB} stands for \emph{Not is this Band} representing that the attenuation is in another band for the class $c = 6$.
The results are shown for different frequency bands: $n=0$ (a), $n=1$ (b), $n=2$ (c), $n=3$ (d), and $n=4$ (e).\\

%{8.5}{9.54}  {7}{7.86}  {4.45}{5}
\insertfig{htbp}{fig:CM}{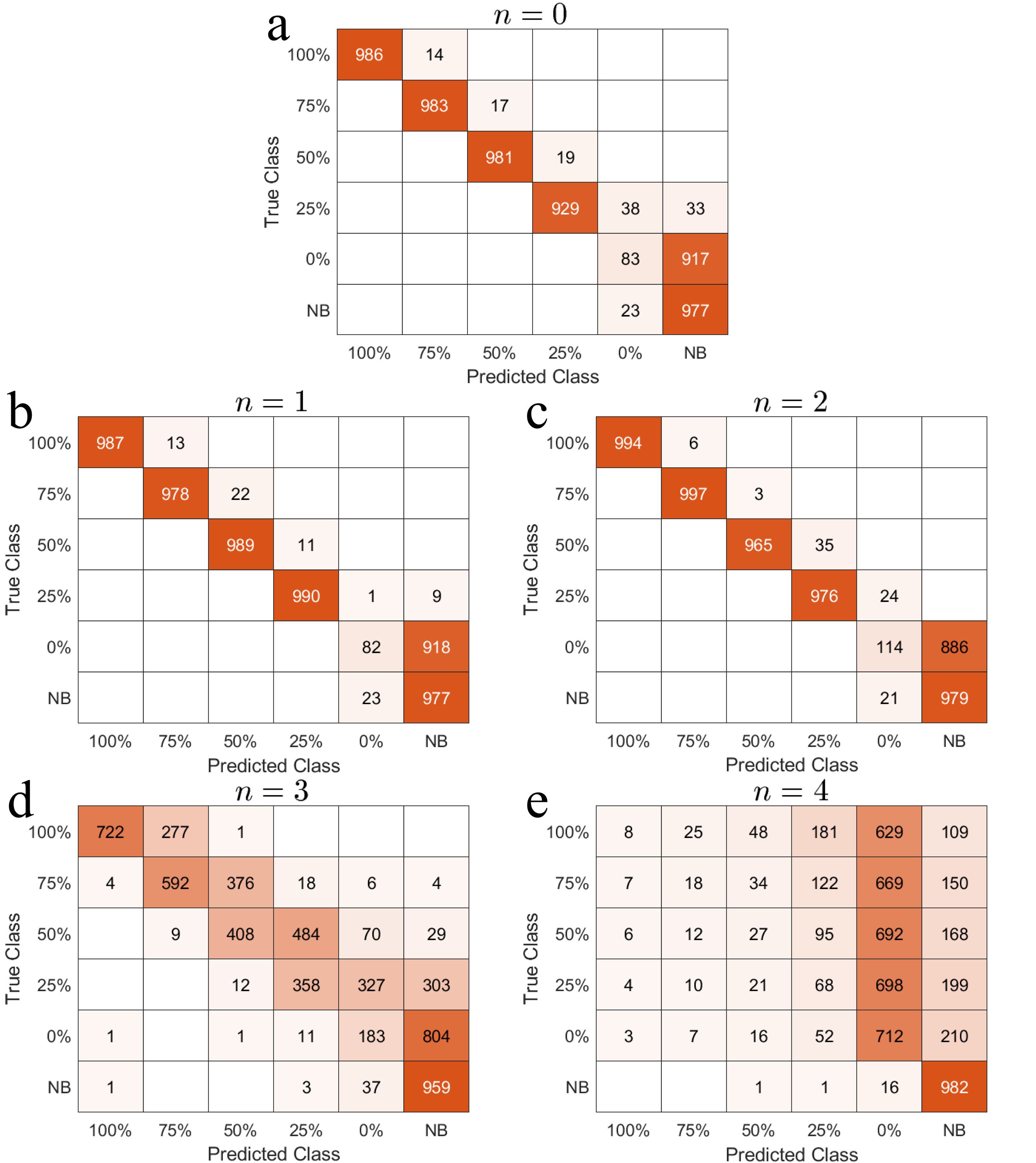}{8.5}{9.54}{Average confusion matrices (50 matrices averaged) for aw-sets in the ${\rm sinc}$-pulse example (1000 elements per output class). The labels denote $\Delta^{(c)}_a$ for the true and predicted class ($c = \{1,2,3,4,5\}$); the label {\rm NB} (Not in this Band) represents $c = 6$. (a,b,c,d, and e): results for different bands.}
%{Average confusion matrices (50 matrices averaged) for aw-sets in the ${\rm sinc}$-pulse spectral attenuation example with 1000 elements per output class. The labels denote de attenuation factor $\Delta^{(c)}_a$ for the true and predicted class with $c = \{1,2,3,4,5\}$; the label {\rm NB} (Not in this Band) represents the class $c = 6$. It is shown the results for the frequency band with $n=0$ (a), $n=1$ (b), $n=2$ (c), $n=3$ (d), and $n=4$ (e).}

An application example of the spectral-correction algorithm is shown in \fig{fig:SC_D1} for a ${\rm sinc}$-pulse waveform $f(t)$ ((b.1), dotted red line).
The waveform $\widetilde{f}(t)$ ((a.1), solid red line) is a ${\rm sinc}$ pulse deformed by a particular spectral response ${\rm F_S}(\nu)$ of a physical system ((a.2), dotted light green line); thus, $\widetilde{f}(t)$ is the output signal of the system with $f(t)$ as the input one.
The spectrum of $\widetilde{f}(t)$ is ${\rm \widetilde{F}}(\nu)$ ((a.2), solid red line), which is obtained by multiplying the spectrum ${\rm F}(\nu)$ of the ${\rm sinc}$ pulse ((b.2), dotted red line) times the response ${\rm F_S}(\nu)$.
The corrected signal obtained by applying the algorithm in \fig{fig:alg_Correct} is $\widehat{f}(t)$ ((a.1), dashed cyan line), whose spectrum is ${\rm \widehat{F}}(\nu)$ ((a.2), dashed cyan line).
The outcome of the correction factors is
\begin{gather*}
\overline{\Delta}_{a,0} = 0.5023,\quad \overline{\Delta}_{a,1} = 0.1259,\quad \overline{\Delta}_{a,2} = 0,\\
\overline{\Delta}_{a,3} = 0,\quad \overline{\Delta}_{a,4} = 0.0001.
\end{gather*}
The corrected-signal spectrum ${\rm \widehat{F}}(\nu)$ can be obtained by multiplying ${\rm F}(\nu)$ times the spectral compensation ${\rm F_C}(\nu)$ ((a.2), dash-dot blue line), where
\[{\rm F_C}(\nu) \equiv \prod_n \left(1 - \overline{\Delta}_{a,n}\,\fnFC{n}(\nu)\right)^{-1}.\]
The signal $f^{*}(t)$ ((b.1), solid light green line) is the output of the system by applying $\widehat{f}(t)$ as the input signal, whose spectrum is ${\rm F}^{*}(\nu)$ ((b.2), solid light green line).
We can conclude about the quality of the compensation result by comparing the waveforms and spectra in (b.1) and (b.2). Besides, it is possible to define an RMS error of the output waveform as
\begin{equation}\label{eq:err_SC}
e_{{\rm RMS\%}} \equiv \frac{100}{max{\abs{f}}}\,\sqrt{\frac{1}{N_{\!s}}\sum_{i = 1}^{N_{\!s}} \left(f^{*}(t_i) - f(t_i)\right)^2},
\end{equation}
where the index $i$ runs over the number of signal samples taken in the time $t_i$ ($N_{\!s}$ samples), and $max{\abs{f}}$ is the maximum absolute value of the signal.
In our example, it is obtained $e_{{\rm RMS\%}} = 4.63\%$.\\

%{8.5}{4.78}  {7}{3.94}  {5.5}{3.1}
\insertfig{htbp}{fig:SC_D1}{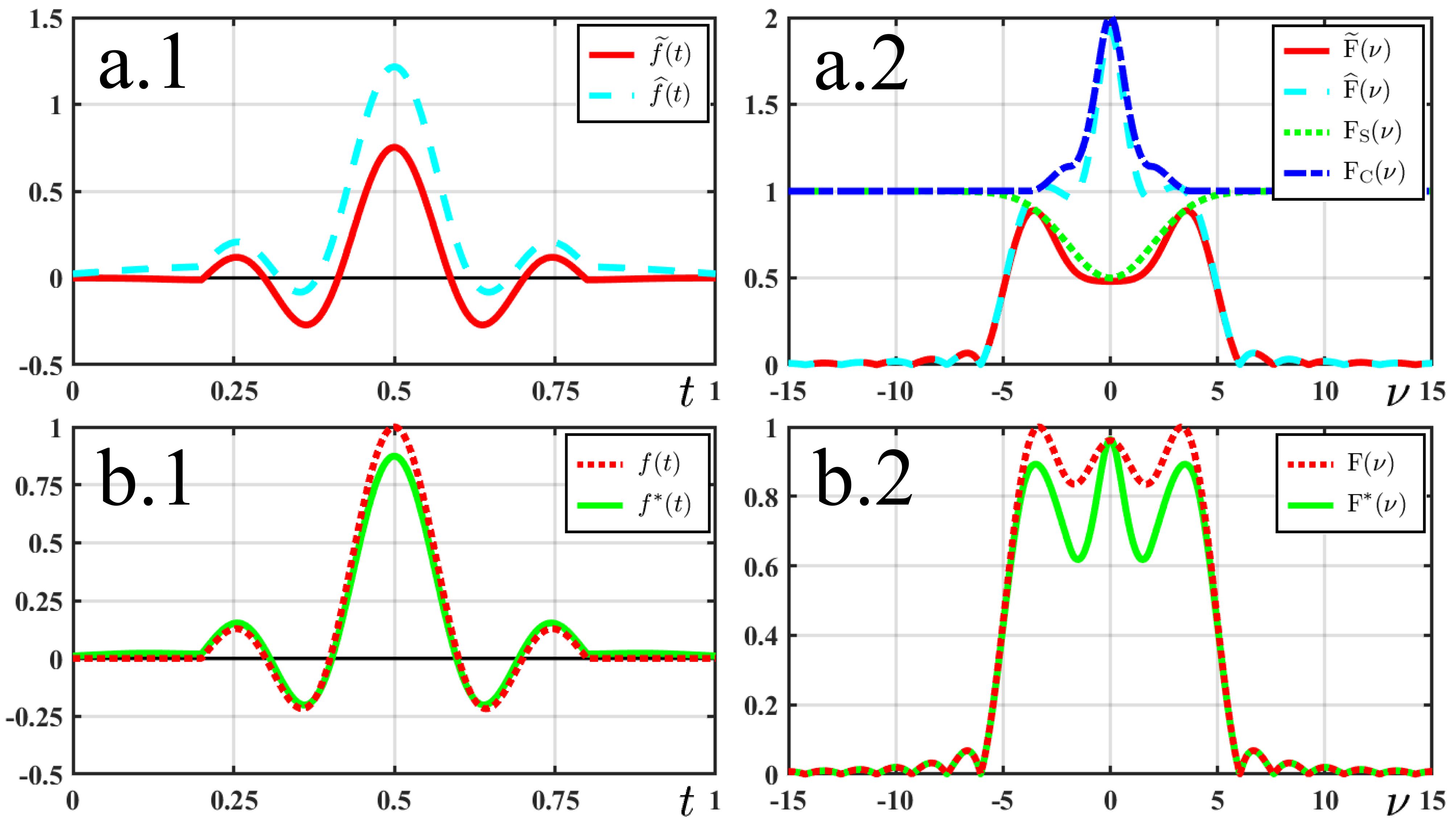}{8.5}{4.78}{Example of spectral-correction algorithm application for a ${\rm sinc}$-pulse waveform. (a.1): $\widetilde{f}(t)$ (solid red line) is the output signal when the original waveform ($f(t)$) is affected by the physical system, $\widehat{f}(t)$ (dashed cyan line) is the input signal corrected by the algorithm; (a.2): ${\rm \widetilde{F}}(\nu)$ (solid red line) is the spectrum of $\widetilde{f}(t)$, ${\rm \widehat{F}}(\nu)$ (dashed cyan line) is the spectrum of $\widehat{f}(t)$, ${\rm F_S}(\nu)$ (dotted light green line) is the spectral response of the system, ${\rm F_C}(\nu)$ (dash-dot blue line) is the spectral correction of the input signal $f(t)$; (b.1): $f(t)$ (dotted red line) is the desired output signal or the original waveform, $f^{*}(t)$ (solid light green line) is the output signal corrected by the algorithm, (b.2): ${\rm F}(\nu)$ (dotted red line) is the spectrum of $f(t)$, ${\rm F}^{*}(\nu)$ (solid light green line) is the spectrum of $f^{*}(t)$.}

In the following section, we discuss the obtained results and arrive at conclusions about the proposed spectral-correction methodology.

%\[\widetilde{f}(t) \quad \widehat{f}(t) \quad f^{*}(t) \quad f(t)\]
%\[{\rm \widetilde{F}}(\nu) \quad {\rm \widehat{F}}(\nu) \quad {\rm F}^{*}(\nu) \quad {\rm F}(\nu)\]
%\[{\rm F_S}(\nu) \qquad {\rm F_C}(\nu) \qquad \prod_n \left(1 - \overline{\Delta}_{a,n}\,\fnFC{n}(\nu)\right)^{-1}\]

\section{Discussion and conclusion}\label{sec:disc_conc}

The proposed time-domain method to correct spectral amplitude attenuations is based on discriminating between a set of different known altered waveforms (aw-set) by using ESSC to extract signal features and simple ANNs to realize the pattern recognition task. Then, a corrected signal is generated as the system input by employing a wavelet decomposition weighted with a series of predicted mean correction factors. Therefore, the physical system alters the corrected signal, and the output will be the desired waveform. Consequently, the precision of the correction algorithm is associated with: the quality of the known aw-set to describe the different spectral deformations produced by the physical system, the capacity of the ESSC routine, and the trained ANNs, to identify the altered waveforms.\\

We observe from \fig{fig:CM}, comparing the predicted class output with the known true class, that in the lower bands ($n = \{0,1,2\}$, (a), (b), and (c)), which hold the most relevant part of the signal spectrum, there is very high effectiveness in recognizing between different waveforms from the 4-factors attenuation group: 100\%, 75\%, 50\%, and 25\% ($c = \{1,2,3,4\}$), that is reflected by the high values of the principal diagonal.
It exists for a particular factor a very-subtle influence of the immediately lower one, i.e., there are elements in the first diagonal superior with low values. Besides, such a very-high performance is also obtained by discriminating between the 4-factors group and the class 0\% ($c = 5$, the original waveform) and {\rm NB} ($c = 5$).
We can conclude that the excellent performance in distinguishing the waveforms is product of generating an aw-set with subtle differences between its members with nearby classes, as it is observed from \figs{fig:Sinc_C0}, \numfig{fig:Sinc_C1}, and \numfig{fig:Sinc_C2}, but distinguishable by the ESSC method and the training of the ANNs. However, there is interference between the outcome of the classes 0\% and NB. That is due to the NB class ($c = 6$) includes waveforms from the aw-set with $n = 4$, whose members are very similar between them and with the original signal (0\%, $c = 5$). Nevertheless, such interference does not affect the calculation of mean correction factors $\overline{\Delta}_{a,n}$ because $\Delta^{(6)}_a = \Delta^{(5)}_a = 0$  (see \eq{eq:Prom_Delta_an}). Therefore, we can adequately correct the altered signals in the frequency bands with $n = \{0,1,2\}$ by efficiently recognizing the 4-factors group waveforms in the aw-set, where the signal spectrum has a significant amplitude (see \fig{fig:DE_Sinc} (b.2), (c.2), and (d.2)).\\
In the case of the band $n = 3$, the attenuation factors are strongly influenced by the immediately lower one; this is observed by the same-order values of the principal diagonal and the first diagonal superior elements in (d).
That is because an altered-waveform class has a very similar form to the waveform of its nearby lower class for this band, as can be seen in \fig{fig:Sinc_C3}.
Therefore, we conclude that the mean correction factor of this band would have an intermediate value between the classes associated with such elements. However, this factor has a poor influence on the corrected signal because of the small-signal spectrum amplitude in the band (see \fig{fig:DE_Sinc} (e.2)).\\
Finally, the band with $n = 4$ holds the higher frequency region, where the signal spectrum has a very low amplitude. Thus, some attenuation in this band does not significantly change the form of the original signal, as we see from \fig{fig:Sinc_C4}, and the predicted class results are attenuations of 0\%, as can be seen in the matrix (e). Therefore, the conclusion is that the mean correction factor of this band would have a value close to 0. Nevertheless, the contribution of this factor to the correction algorithm is despicable because the signal spectrum has a negligible amplitude in this band (see \fig{fig:DE_Sinc} (f.2)).\\

We obtained from the application example shown in \fig{fig:SC_D1} that the spectrum-corrected algorithm outcome is an output-system signal with an RMS error lower than 5\% (see \eq{eq:err_SC}). That performance is achieved with a significantly spectral deformation supposed to the physical system response (see ${\rm F_S}(\nu)$ in \fig{fig:SC_D1} (a.2), dotted light green line). Therefore, we conclude that the algorithm can adequately compensate these spectral-alteration types on the signal.\\

There are pending issues to be analyzed, which are out of the scope of this work, between then: i) test the performance of the spectral-correction method by changing the noise level or signal-to-noise ratio, ii) study the precision of recognizing spectral profiles by changing the width of the band-rejection filters and using different band numbers, iii) use a unique global ANN to detect all possible spectral alterations by combining attenuations in different frequency bands.\\
The approach used in this work is to detect a general spectral-attenuation profile by splitting the spectrum into frequency bands and using one ANN per band. We suppose that combinations of attenuations in different bands can be detected by the join working of all the ANNs. Therefore, each ANN must discriminate between $N_p = n_l + (n_l-1)\,n_{\!M}$ different patterns of altered waveforms, where $n_l$ is the number of attenuation levels; then, $N_p = 21$ in our case ($n_l = 5$, and $n_{\!M} = 4$).
If we use a unique ANN to detect different spectral attenuation, such an ANN is trained using all the possible combinations of attenuations in the bands, but there are many combinations with subtle spectral altered waveforms similar to the original one; thus, there would be more confusion in the detecting procedure. Besides, the number of patterns to discriminate will be $N_p = n_l^{n_{\!M}+1}$; then, $N_p = 3125$ in our case.\\

As a final remark, the results obtained from the confusion matrices and the application example of \sect{sec:ex_sac_sinc} permit us to conclude that a spectral-correction procedure can be satisfactorily performed by the proposed time-domain algorithms (using ESSC and simple ANNs). Thus, this method is convenient for embedded or limited computer architectures.\\

%\[n=0 \quad n=1 \quad n=2 \quad n=3 \quad n=4\]

\section*{Acknowledgment}

This work was supported by CONICET and FONCYT (PICT 2013-2600), Agencia Nacional de Promoción de la Investigación, el Desarrollo Tecnológico y la Innovación (MINCYT).

%\bibliographystyle{unsrt}
%%\bibliography{ref_SA-TDANN}
%\bibliography{ref_SA-TDANN_title}

\end{document}